\documentclass[twocolumn,showpacs,preprintnumbers,amsmath,amssymb]{revtex4}

\newcommand{\be}{\begin{eqnarray}}
\newcommand{\ee}{\end{eqnarray}}
\newcommand{\etal}{{\em et al} }

\usepackage{graphicx}
\usepackage{dcolumn}
\usepackage{bm}

\begin{document}

\preprint{DRAFT}

\title{Fully Correlated Electronic Dynamics for Antiproton Impact Ionization of Helium}

\author{M. Foster}
\author{J. Colgan}%
\affiliation{%
Theoretical Division, Los Alamos National Laboratory, Los Alamos, New Mexico 87545, USA\\}%

\author{M. S. Pindzola}
\affiliation{Department of Physics, Auburn University, Auburn, Alabama 36849, USA}

\date{\today}

\begin{abstract}
We present total cross sections for single and double ionization of helium by antiproton impact over a wide
range of impact energies from 10 keV/amu to 1 MeV/amu.
A non-perturbative time-dependent close-coupling method (TDCC) is applied to fully treat the  
correlated dynamics of the ionized electrons. Excellent 
agreement is obtained between our calculations and experimental measurements of total single and double 
ionization cross sections at high impact energies, whereas for lower impact energies, some discrepancies
with experiment are found. At an impact energy of 1 MeV we also find that the double-to-single ionization
ratio is twice as large for antiproton impact as for proton impact, confirming a long-standing unexpected
experimental measurement.
\end{abstract}

\maketitle

The double ionization of helium by ion impact has been a long and fruitful field of study in atomic collision
physics. As a fundamental four-body system, it provides stringent tests of any theoretical description of 
charged particles moving in a Coulomb field. 
Moreover, the sign and magnitude of the ion charge can be experimentally varied relatively simply 
to explore interesting physical effects. 
New antimatter collision experiments are planned at FAIR, 
the Facility for Antiproton and Ion Research \cite{flair06}.  
FAIR is an international collaboration on atomic and molecular physics that intends to 
investigate antiproton driven ionization processes and even kinematically complete antiproton collision experiments.
These experimental efforts complement the recent intense activity in antihydrogen studies of recent years \cite{gab1}.
In the last 20 years, experimental measurements showed, unexpectedly, that the ratio of double-to-single
ionization of helium from proton impact was around a factor of two lower than that from antiproton impact
\cite{andersen86}, a feature which has not yet been observed in a fully converged theoretical calculation.
Subsequent  experimental measurements \cite{hvelplund94} of the total double ionization cross section
by antiproton impact not only revealed a larger cross section than for proton impact, but also that at low
impact energies, the double ionization cross section for antiproton impact does {\it not} monotonically
decrease, but shows an increasing cross section as the impact energy is lowered. 

In response to this renewed experimental activity, 
in this Letter we present fully converged calculations of single and double ionization of helium from antiproton impact
and find similar 
unexpected ratios of the double-to-single ionization of helium compared with proton impact.
We also provide converged single and double ionization cross sections for antiproton impact over a wide
energy range from 10 keV to 1 MeV.
Theoretical calculations for single ionization of helium by antiproton impact cross sections have 
suggested that the experimental measurements \cite{hvelplund94} below 30 keV may be inaccurate 
\cite{schultz03, lee00}.  Experimental efforts are underway to remeasure the single and 
double ionization cross sections \cite{ichioka05} and preliminary results \cite{knudsen07} suggest 
that the single ionization cross section does not decrease as rapidly as reported by Hvelplund \etal 
\cite{hvelplund94}.  These preliminary results may bring closer agreement between theory and experiment for single
ionization.  

For double ionization of helium by antiproton impact, no theoretical description has been used to calculate
the cross section over the full energy range of the measurements.   
Barna \etal \cite{barna03} calculated the double 
ionization cross section at a single energy of 3.6 MeV which agreed
well with experiment. 
A close-coupling method was used by D{\'i}az \etal \cite{diaz02} to 
calculate double ionization cross sections for impact energies above 200 keV, with limited success. The calculations
reported were around a factor of two lower than the measurements of \cite{hvelplund94}.
Keim \etal calculated the single and double ionization of helium by antiproton impact using the framework 
of time-dependent density-functional theory \cite{keim03}.  
These results produced reasonable agreement for the single ionization 
cross sections.  However, the double ionization cross sections 
were considerably higher than the experimental results for all energies.   
A similar approach was used by Tong \etal \cite{tong02}, who calculated 
single ionization cross sections that were lower than the experimental measurements, 
and double ionization probabilities for 15 keV and 100 keV that 
yield much larger cross sections compared with experiment.  
The multicut forced impulse 
method (mFIM) has been the only calculation for double ionization of helium by antiproton impact below 
200 keV \cite{bronk98} which has compared well with experiment. However, at these low impact energies
the mFIM method predicts single ionization cross sections that are in less than satisfactory agreement with previous calculations \cite{schultz03,lee00}.
  

In this work, we apply our TDCC method to treat the double ionization 
of helium by antiproton impact \cite{pindzola07}. 
The six-dimensional time-dependent Schr{\"o}dinger equation for the outgoing electrons is reduced
to a coupled set of two-dimensional partial differential equations through a partial-wave expansion of the
two-electron wavefunction in spherical coordinates centered  on the target atom.
The two-electron wavefunction is subjected to 
a time-dependent projectile interaction which is included through a multipole expansion. We assume 
a straight-line motion for the impacting ion.  

Our previous study using this approach explored alpha-particle collisions with helium. For high impact 
energies, excellent agreement between theory and experiment was found for 
single and double ionization of helium. At low energies, charge transfer to the impacting ion may become
significant, so that a description using a two-electron wavefunction centered on the target is unsuitable.
However, in the case of antiproton impact, charge transfer cannot occur, so our method is appropriate for 
treatment of the antiproton-atom collision at any impact energy.

The fully correlated wavefunction, $\Psi^{LM}$,
for the single and double ionization of a two-electron target atom
by antiproton collision is obtained by the evolution of
the time-dependent Schr\"odinger equation in real time:
\begin{equation}
 i{\partial\Psi^{LM}(\vec{r_1},\vec{r_2},t)\over\partial t}
  = H_{system} \Psi^{LM}(\vec{r_1},\vec{r_2},t) \ ,
\end{equation}
where the non-relativistic Hamiltonian is given by
\begin{eqnarray}
H_{system} =&& \sum^2_i(-{1\over 2}\nabla^2_i - {Z_t\over r_i})+{1\over |\vec{r_1}-\vec{r_2}|}
\nonumber\\
&&-
{Z_p\over |\vec{r_1}-\vec{R}(t)|} - {Z_p\over |\vec{r_2}-\vec{R}(t)|}
\end{eqnarray}
and $Z_p$ is the projectile atomic number and $Z_t$ is the target atomic number.
For straight-line motion,
the magnitude of the time-dependent projectile position is given by
\begin{equation}
 R(t) = \sqrt{b^2 + (d_0 + vt)^2} \ ,
\end{equation}
where $b$ is an impact parameter, $d_0$ is a starting distance ($d_0 < 0$), and
$v$ is the projectile speed.
If we expand $\Psi^{LM}$ in coupled spherical harmonics and substitute
into Eq.~(1), the resulting close-coupled equations for the
 radial expansion functions, $P^{LM}_{l_1l_2}(r_1,r_2,t)$, are then given by
\begin{widetext}
\begin{eqnarray}
  i{\partial P^{LM}_{l_1l_2}(r_1,r_2,t)\over\partial t}
  & = &
  T_{l_1l_2}(r_1,r_2) P^{LM}_{l_1l_2}(r_1,r_2,t)
  +\sum_{l'_1,l'_2} V^{L}_{l_1l_2,l'_1l'_2}(r_1,r_2)
                    P^{LM}_{l'_1l'_2}(r_1,r_2,t)
  \\
  &+&
  \sum_{L'M'}\sum_{l'_1,l'_2} W^{LM,L'M'}_{l_1l_2,l'_1l'_2}(r_1,R(t))
                    P^{L'M'}_{l'_1l'_2}(r_1,r_2,t)  +\sum_{L'M'}\sum_{l'_1,l'_2} W^{LM,L'M'}_{l_1l_2,l'_1l'_2}(r_2,R(t))
                    P^{L'M'}_{l'_1l'_2}(r_1,r_2,t) \ ,\nonumber 
\end{eqnarray}
\end{widetext}
where $V^{L}_{l_1l_2,l'_1l'_2}(r_1,r_2)$ is the electron-electron potential and 
$W^{LM,L'M'}_{l_1l_2,l'_1l'_2}(r_i,R(t))$ represents the interaction between the projectile and the electrons.  The initial value boundary condition for Eq. (4) is given by 
\begin{equation}
 P^{LM}_{l_1l_2}(r_1,r_2,t=0)
 = \delta_{L,L_0} \delta_{M,M_0}
   \bar{P}^{L_0M_0}_{l_1l_2}(r_1,r_2,\tau\rightarrow\infty) \ ,
\end{equation}
where $\bar{P}^{L_0M_0}_{l_1l_2}$ is the radial portion of the ground state wavefunction.
The two-electron ground state wavefunction of the helium atom is obtained by relaxation of the time-dependent Schr\"{o}dinger equation in imaginary time ($\tau = $i$t$).  The fully correlated ground state is expanded in coupled spherical harmonics resulting in a set of close-coupled equations for the $\bar{P}^{L_0M_0}_{l_1l_2}(r_1,r_2,\tau)$ radial expansion functions with the initial boundary conditions given by a product of single particle bound radial orbitals for the one-electron target ion with $L=M=0$ \cite{pindzola07}.

We solve the TDCC equations, Eq. (4), using a lattice technique to obtain a discrete 
representation of the radial expansion function and all operators on a two-dimensional grid.  
The single and double ionization probabilities, ${\cal P}(E,b)$, are calculated using the 
time-dependent radial wavefunction propagated for a suitable time. The radial wavefunction is then projected onto both bound and continuum single particle orbitals, as discussed in detail in \cite{pindzola07}.

The TDCC calculations employed a 384 x 384 point radial lattice with a uniform mesh spacing of $\Delta r = 0.20$, 
giving a converged ground state of helium on the lattice. In order to achieve accurate double 
ionization cross sections it was found that up to 101 coupled channels had to be included in 
the expansion of the final state two-electron wavefunction.  
The time-dependent projectile potential was expanded into monopole, dipole, quadrupole, and octopole terms, 
all of which were required to converge the total cross sections.  The time-dependent wavefunction was 
propagated from a time when the projectile was a distance $d_0 = -50 a_0$, through closest approach, to a 
time when the projectile was at a distance $d_0 = +40 a_0$.  The total cross sections calculated at the lowest
impact energy of 10 keV 
had to be propagated for $142 a.u.$ of time to reach this distance using a number of time steps greater 
than 28,000.  \linebreak 

\begin{figure}[!ht]
\label{fig1}
\begin{center}
\includegraphics[scale=0.28]{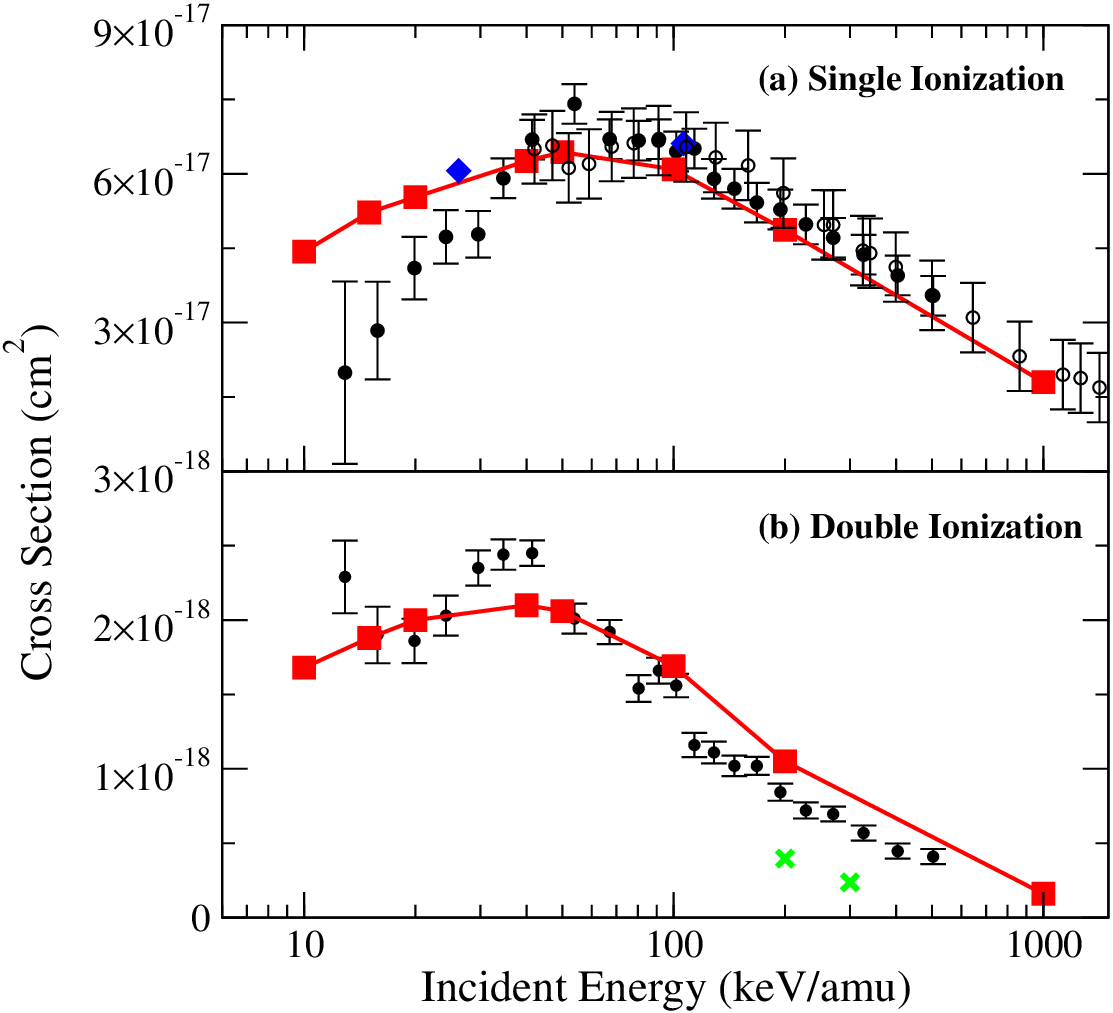}
\caption{(a) Cross sections for single ionization of helium by antiproton impact.
Filled squares: TDCC calculations; filled diamonds: calculations of Schultz
and Krstic \cite{schultz03}; filled and open circles: experimental measurements of \cite{hvelplund94, andersen90}.
(b) Cross sections for double ionization of helium by antiproton 
impact. Filled squares: TDCC calculations; crosses: calculations of Diaz 
{\em et al}\cite{diaz02}; filled circles: experimental measurements of \cite{hvelplund94}.}
\end{center}
\end{figure}

In Fig.~1, we present our TDCC calculations for single and double ionization of helium by antiproton collisions.
The top panel of Fig.~1 shows the single ionization cross section measurements \cite{hvelplund94,andersen90}, 
our TDCC results (red squares), and the calculations of Schultz and Krstic \cite{schultz03}. The two sets of 
calculations are in good agreement over a wide energy range down to around 30 keV. Below 30 keV, 
the TDCC calculations are higher than the experimental measurements, although new preliminary measurements of
this single ionization cross section \cite{knudsen07} suggest that the previous measurements \cite{hvelplund94}
may be too low.

The lower panel of Fig.~1 shows the double ionization of helium by antiproton impact for ion energies from 
10 keV to 1 MeV.  
The TDCC (red squares) calculations are in excellent agreement with the experimental measurements above 20 keV. 
Between 10 and 20 keV, the TDCC calculations decrease monotonically. 
However, the experimental measurements show a sharp increase as the impact energy decreases.
The TDCC results are in excellent agreement with the experimental data points at 15 keV and 20 keV, 
suggesting that perhaps only the lowest energy measurements are in error.
\linebreak

\begin{figure}[!ht]
\label{fig2}
\begin{center}
\includegraphics[scale=0.28]{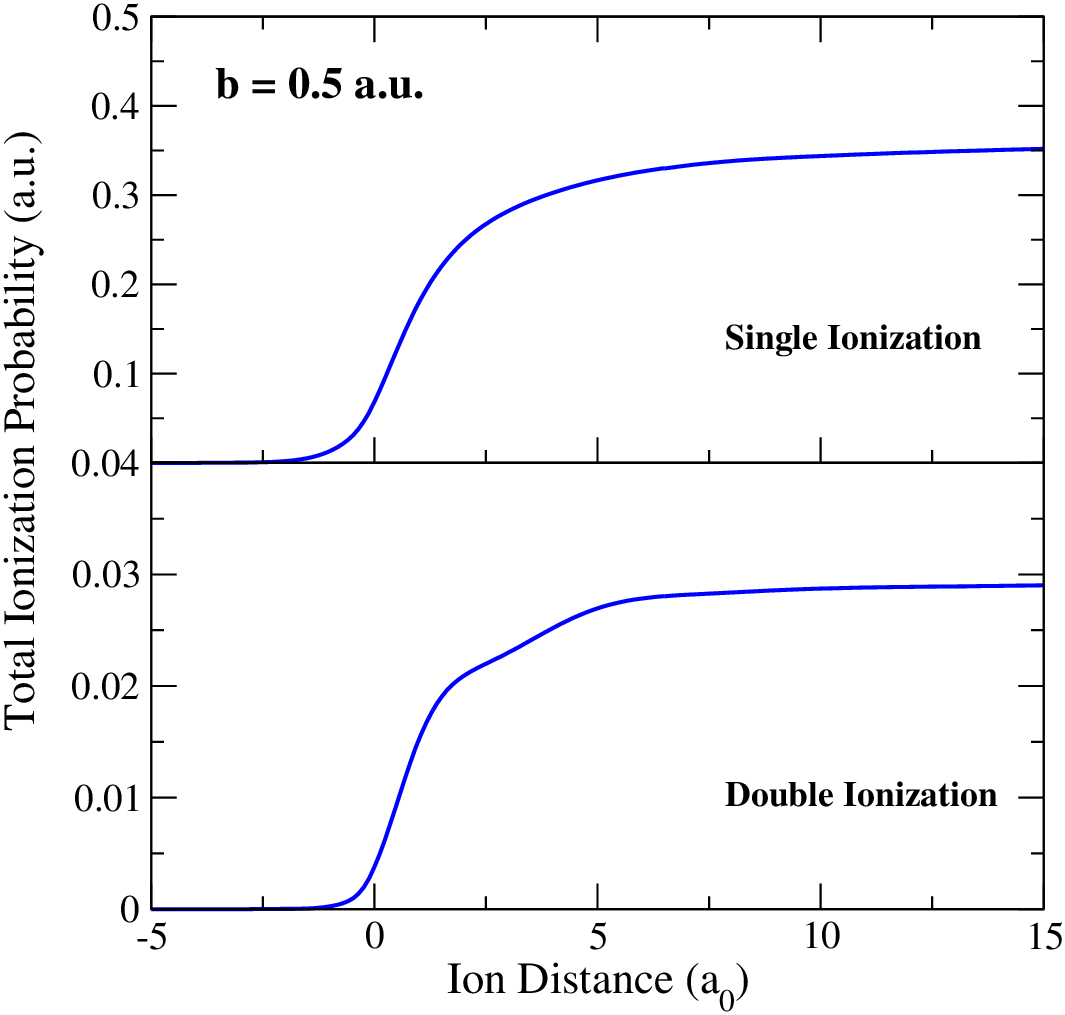}
\caption{Evolution of the ionization probability, ${\cal P}(E,b = 0.5 a_0)$, for a 50 keV antiproton 
collision with a helium atom as a function of the impacting ion distance.
Upper panel: single ionization probability summed over all partial waves; lower panel: double ionization probability. 
The helium atom is located at the origin of 
the collision system ($d_0=0$). }
\end{center}
\end{figure}

A time-dependent propagation of the two-electron wavefunction can be a powerful tool in understanding the 
evolution of the collision system.  Fig.~2 shows the ionization probability, ${\cal P}(E,b)$ for both 
single and double ionization for a 50 keV antiproton collision with a helium atom.  The initial ion 
starts at $d_0 = -50 a_0$, then propagates through distance of closest approach ($b = 0.5 a_0$) to a 
final distance of $d_0 = +40 a_0$.  The total propagation time for the 50 keV collision system is $63 a.u.$ 
of time.  The helium atom is placed at $d_0 = 0 a_0$.  Fig.~2 shows the evolution of the ionization probability for one impact parameter, 
($b = 0.5 a_0$) and for an ion distance from $d_0 = -5 a_0 \rightarrow 15 a_0$.   The choice of the impact parameter equal to $b = 0.5 a_0$ 
represents the impact distance of maximum double ionization probability.  The total single and double ionization 
cross sections for a given energy are obtained by the relation
\begin{equation}
 \sigma(E) = 2\pi \int^{\infty}_0 {\cal P}(E,b) b db \ .
\end{equation}
The ionization probabilities for both single and double ionization of helium by antiproton impact 
increase rapidly just before the antiproton reaches the helium atom (positioned at $d_0 = 0 a_0$),
and then tend to a constant value beyond $d_0 \sim 10 a_0$.  The results in Fig.~2 show 
that the antiproton collision with the helium atom occurs in approximately $t_{collision} = 4.24$ a.u. of time 
or $0.1$ fs.  Such fast collision times are currently just beyond the pulse lengths of 
experimental ultrafast laser techniques which are being considered for use in probing
electronic behavior of atomic systems.  
\linebreak

\begin{figure}[!ht]
\label{fig3}
\begin{center}
\includegraphics[scale=0.28]{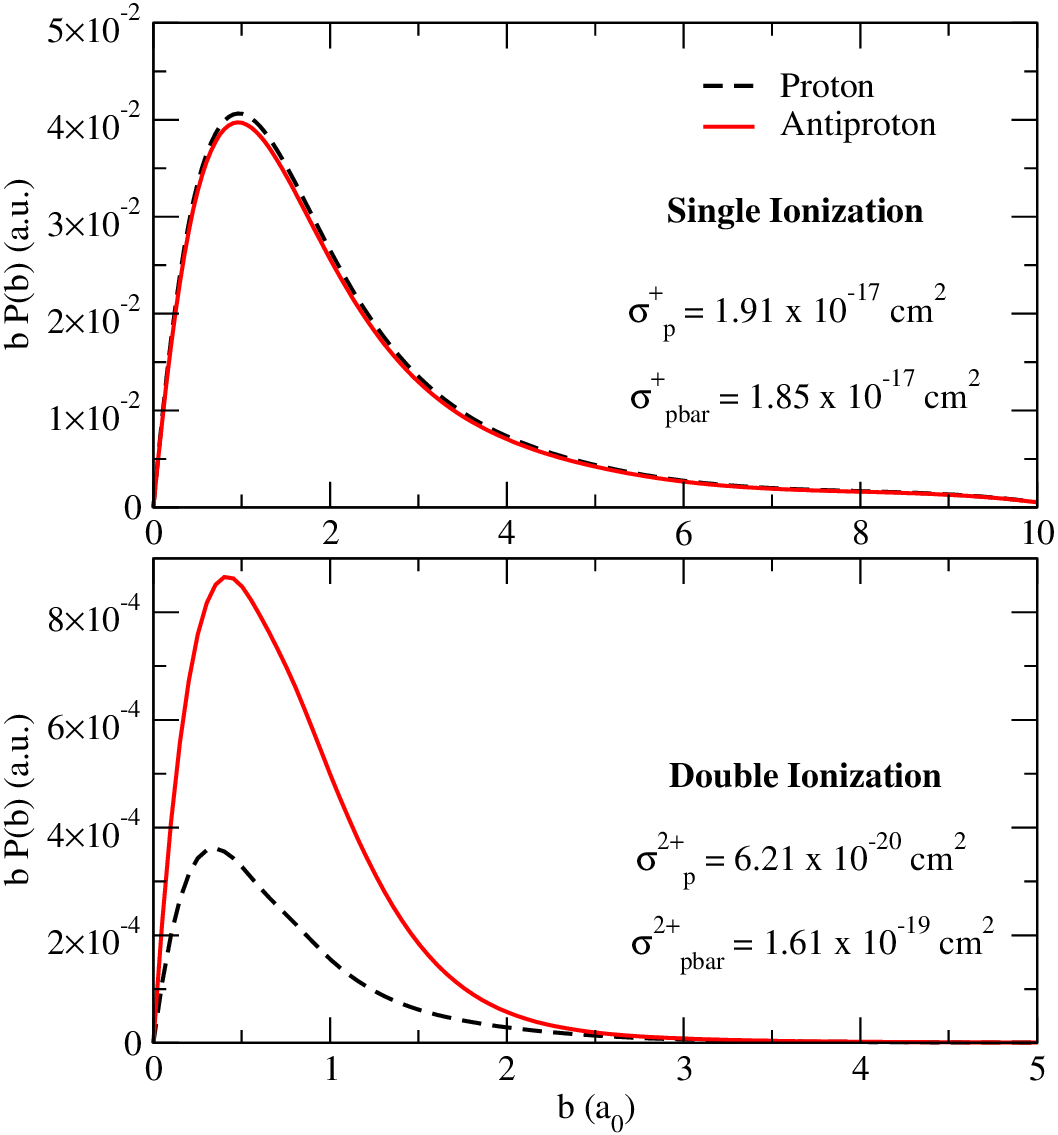}
\caption{Weighted probabilities for single (upper) and double (lower) ionization of helium by antiproton (red line) and proton (black dashed line) impact  at 1 MeV impact energy.}
\end{center}
\end{figure}

The total ionization cross sections are calculated by integrating the ionization probabilities as defined in Eq.~(6).  
Fig.~3 shows the weighted probabilities (the integrand of Eq.~(6)) for both proton (black dashed line) and 
antiproton (red solid line) ionization as a function of the impact distance from 
the helium atom at an incident energy of 1 MeV. For the antiproton impact, the peak of the single ionization weighted probability is 
at an impact parameter of 
$b \approx 1.0 a_0$, while the double ionization weighted probability peaks at a smaller impact parameter ($b \approx 0.8 a_0$).  
The proton's maximum ionization probability is similar to the antiproton for single ionization but shifted to smaller impact parameters 
for double ionization.  However, of much greater significance is the difference between the proton and antiproton double 
ionization probabilities.  As also observed in the experimentally measured double-to-single ratios \cite{andersen86, hvelplund94}, 
our calculations find that for antiprotons the double ionization cross section is larger than the proton double ionization by 
approximately a factor of 2, but that the single ionization processes have approximately the same total cross section.  
At this impact energy, the total single and double ionization cross sections are in good agreement with experiment
for antiproton impact (as shown in Fig.~1) and for proton impact (compared with the measurements of \cite{shah85}).
The reasons for this difference between the double ionization cross sections is still not well understood. Clearly, the change of
sign of the projectile affects the double ionization mechanism in a much greater manner than the single ionization mechanism. 
This could be partially explained by arguments arising from a Born-approximation description \cite{mcguire1},
where interference between various mechanisms for double ionization depended on the sign of the ion
charge, resulting in different double-to-single ratios for protons and antiprotons. 
We note that following a rapid ejection of a single electron, which based on the upper panel of figure 3 is equally probable for the proton impact and antiproton impact, the remaining transient collision complexes are quite different. For proton impact, the remaining electron sees a positive charge of +3, while for the anti-proton impact the remaining electron sees a positive charge of +1.  Thus, the second electron may find it easier to escape in the antiproton impact case.  Similar effects have been observed for double photoionization of helium by large photon energies, where the shake-off mechanism becomes dominant at high photon energy \cite{knapp2002, kheifets2001}.  Although complicated by the multipole potential is this case, a similar mechanism may be responsible for the large double ionization probability from the antiproton impact.  Also, the impact distance, $b$, can provide useful insight into 
understanding the dynamics involved in the collision. 
Clearly, double ionization is most probable when the antiproton significantly penetrates the electron cloud of the helium atom. This suggests the 
possibility  that the antiproton undergoes multiple collisions with the electrons to doubly ionize. 
Single ionization is most probable at a somewhat
larger impact parameter, and still is probable even when the ion is three or four atomic units from the atom, at which distance there is 
almost no probability of double ionization. Thus the single ionization mechanism 
is likely dominated by a more standard `binary' collision process.


In conclusion, the TDCC method has been shown to accurately compute single and double ionization total cross sections 
for antiproton impact ionization of helium over a wide energy range. The single ionization total cross section calculations are in excellent
agreement with experiment and with previous calculations. The double ionization total cross section calculations also agree well with
experiment above 20 keV.   
The antiproton double ionization cross sections are 2 times larger than 
the proton double ionization cross sections at the same impact energy
whereas, the single ionization cross sections are similar.
We aim to further exploit our method to extract the differential cross sections for the two outgoing electrons after ion impact. Such calculations
may allow comparison with the large set of experimental measurements of differential cross sections for ion impact which currently exist
(e.g. \cite{khayyat99}), and 
should yield more insight into the nature of the double ionization process.

A portion of this work was performed under the auspices of the US Department of Energy through Los Alamos 
National Laboratory and through DOE and NSF grants to Auburn University.
Computational work was carried out at the National Center for Computational Sciences in Oak Ridge, TN, and
using the Institutional Computing Resources at Los Alamos National Laboratory. \linebreak


\end{document}